\documentstyle[12pt,aaspp4]{article}
\input psfig.sty

\begin{document}

\title{Jet directions in Seyfert galaxies: Radio continuum imaging data}

\author
{H.R. Schmitt\altaffilmark{1,5}, J.S. Ulvestad\altaffilmark{2},
R.R.J. Antonucci\altaffilmark{3} and A.L. Kinney\altaffilmark{4}}
\altaffiltext{1}{Space Telescope Science Institute, 3700 San Martin Drive, 
Baltimore, MD 21218, USA}
\altaffiltext{2}{National Radio Astronomy Observatory, P.O. Box 0,
1003 Lopezville Road, Socorro, NM87801}
\altaffiltext{3}{University of California Santa Barbara, Physics Department,
Santa Barbara, CA93106}
\altaffiltext{4}{NASA Headquarters, 300 E St., Washington, DC20546}
\altaffiltext{5}{email:schmitt@stsci.edu}

\date{\today}

\begin{abstract}

We present the results of VLA A-array 8.46 GHz continuum imaging
of 55 Seyfert galaxies (19 Seyfert 1's and 36 Seyfert 2's). These
galaxies are part of a larger sample of 88 Seyfert galaxies, selected
from mostly isotropic properties, the flux at 60$\mu$m and warm
infrared 25$\mu$m to 60$\mu$m colors. These images are used to study
the structure of the radio continuum emission of these galaxies and their
position angles, in the case of extended sources. These data, combined
with information from broad-band B and I observations, have been used to
study the orientation of radio jets relative to the plane of their host
galaxies (Kinney et al. 2000).

\end{abstract}


\section {Introduction}

Recently Schmitt et al. (1997), Clarke, Kinney \& Pringle (1998) and
Nagar \& Wilson (1999) showed that there is little or no correlation between
the position angle of radio jets and galaxy disk major axes in Seyfert
galaxies.  Employing a statistical inversion technique, which uses the
observed values of $\delta$ (the difference between the position angle
of the jet and the host galaxy disk major axis) and $i$ (inclination of
the galaxy disk relative to the line of sight), Clarke et al. (1998)
and Nagar \& Wilson (1999) showed that the observed
$\delta$-distribution can be reproduced by a homogeneous distribution
of angles $\beta$ (the angle between the jet and the galaxy disk
axis).

These results confirm previous ones based on smaller samples (Ulvestad
\& Wilson 1984b; Brindle et al. 1990; Baum et al. 1993), and contradict
the expectation that the jets should be aligned perpendicular to the
galaxy disk. Assuming that the accretion disk and black hole are fed by
gas from the host galaxy disk, it is natural to expect both disks to be
aligned and have the same angular momentum vector. As a result, since
jets are emitted perpendicular to the accretion disk, we would expect
them to be aligned with the host galaxy minor axis, which is not
observed. These studies give us information about the circumnuclear
region of Seyferts and may be important in the study of processes
involved in the feeding of the AGN. Possible explanations for the
misalignment of the jet and the host galaxy disk are warping of the
accretion disk, which can be caused by the self induced radiation
instability (Pringle 1996; 1997; Maloney, Begelman \& Pringle 1996), or
by the misaligned inflow of gas from minor mergers (Barnes \& Hernquist
1992, 1996). Kinney et al. (2000) give a complete list of possible causes
of which we are aware for the observed misalignment.

The results from Clarke et al. (1998) and Nagar \& Wilson (1999) were
statistically significant. However, both papers were limited by the
fact that they did not use well defined samples and most of their
measurements were obtained from the literature. As a result, their
results could suffer from selection effects, like the preferential
selection of galaxies which have jets shining into the plane of the
galaxy, resulting in brighter radio emission and narrow line regions,
which would be easier to detect. The fact that these papers used data
selected from the literature could also have influenced the results,
because different authors are likely to measure the data using different
techniques and use data of different quality.

In order to improve our analysis relative to previous studies, we
obtained radio continuum images at 8.46 GHz (3.6cm), optical broad band
images and spectroscopy for a sample of Seyfert galaxies selected
from a mostly isotropic property, the flux at 60$\mu$m, and warm
infrared colors. In this way we
avoid selection effects and create a homogeneous database, with
measurements done using consistent techniques.

In this paper we present the VLA A-configuration X-band (8.46 GHz)
observations of 55 of the Seyfert galaxies in our sample. The
information about the remaining 20 galaxies for which there were data
available in the literature is presented by Kinney et al. (2000) as are
the statistical results of the position angle measurements. The
broad-band B and I imaging data is presented by Schmitt \& Kinney
(2000). In Section 2 we present the sample. The description of the
observations and reductions is given in Section 3, and the measurements
are presented in Section 4. A summary is given in Section 5.

\section{The Sample}
\label{sample}

As pointed out in the Introduction, previous works in this field
probably suffered from selection effects, since they used samples
obtained from the literature. In order to avoid these problems as much
as possible, we have chosen a sample selected from a mostly isotropic
property, the flux at 60$\mu$m. Pier \& Krolik (1992) models showed
that the circumnuclear torus of Seyferts radiates nearly isotropically at
60$\mu$m.  Since these are the most anisotropic of the published
models, they also are the most conservative.

We have chosen to use the de Grijp et al. (1987, 1992) and Keel et al.
(1994) sample of warm IRAS galaxies, which were selected based on the
quality of the 60$\mu$m flux measurements, Galactic latitude
$|b|>20^{\circ}$, and 25$\mu$m$-60\mu$m color in the range
$-1.5<\alpha(25/60)<0$. This color criteria excluded starburst galaxies
as much as possible, but the galaxies were also observed
spectroscopically to confirm their activity types (de Grijp et al.
1992). Our sample includes all the Seyferts in de Grijp et al.  (1987,
1992) which had redshift z$\leq0.031$, giving a total of 88 galaxies
(29 Seyfert 1's and 59 Seyfert 2's). This distance limit is large
enough to include a substantial number of galaxies to provide good
statistics, but still close enough to allow the resolution of radio
features.  More details about the sample selection are given by Kinney
et al. (2000).

Of the 88 galaxies in our sample, 77 have $\delta>-47^{\circ}$ and can
be observed from the VLA. Here we present our observations of 36 of
these galaxies, and images of 19 other galaxies for which data were
extracted from the VLA archive. For 20 other galaxies we were able to
obtain measurements from the literature, giving a total of 75 Seyfert
galaxies with high resolution 8.46GHz (3.6cm) radio observations.

Table 1 presents the galaxies observed by us, and those for which we
reduced VLA archival data. The measurements for the entire sample of 88
galaxies, including the galaxies for which we obtained radio
observations from the literature, is presented by Kinney et al. (2000)
and Schmitt \& Kinney (2000). The Table gives the identification
numbers of the galaxies in the de Grijp et al. (1987) catalog, their
names, activity types, dates in which the observations were done, the
codes of the VLA proposals from which the data were obtained, the
observational method used to observe the sources (S) and the phase
calibrators (C), the integrated 3.6cm flux densities, the 1-$\sigma$
noise levels, the position angles of the extended radio structures
(PA$_{RAD}$), the sizes of the radio sources (calculated assuming
H$_0=75$ km s$^{-1}$ Mpc$^{-1}$) their morphologies, and in the case of
archival data, the reference where the data have already been
published.

\section{Observations and reductions}
\label{obs}

All the observations presented in this paper were obtained with the VLA
in A-configuration at 8.46 GHz. Of the 55 galaxies presented here, 36
were observed by us (proposal AA226) and the other 19 were obtained
from 7 other projects for which there were data available in the VLA
archive. The dates in which the galaxies were observed and their
respective VLA proposal codes are given in columns 4 and 5,
respectively, of Table 1.

The observations were done in snapshot mode, using 3C286 and/or 3C48 as
primary flux calibrators (depending on the proposal). Most of the
observations were done using J2000.0 coordinates with the exception of
7 galaxies, observed by the projects AB618, AD244, AK350, AK407 and
AP276, which were observed using B1950.0 coordinates and later
converted to J2000.0 with AIPS. The phase calibration was done using
calibrators from the NRAO list, preferentially using A-category
calibrators closer than 10$^{\circ}$ from the galaxies. Most of the
observations were done sandwiching $\sim$10 minutes observations of the
galaxy (S) with short $\sim$2--3~minutes observations of a phase
calibrator (C), repeating the process for 3 to 7 times, which gives a
total on source integration time of 30 to 90 minutes. For 14 galaxies,
observed in the proposals AK394 and AD244, the source was observed only
once, for $\sim$15 minutes, followed by a $\sim$5 minute observation of
the phase calibrator. The observing methods used for each galaxy are
shown in the 6th column of Table 1, where ``CSC...'' means that the
source was sandwiched with observations of the phase calibrator, and
``CS'' that the source and phase calibrator were observed only once.

The reductions were done using AIPS and following standard techniques.
In the case of sources observed in the ``CSC...'' mode, the phase
calibration was done by linearly interpolating between the observations
of the phase calibrator, while for sources observed in the ``CS'' mode,
the data was first boxcar smoothed and then interpolated. The images
were made using uniform weighting, all the sources were
interactively self-calibrated in phase and imaged.

Typical resolutions in our final images were
0.20$^{\prime\prime}-0.25^{\prime\prime}$. Because the VLA
A-configuration is not sensitive to structures with scale sizes larger
than about 7$^{\prime\prime}$ at 3.6 cm, the results reported here
generally refer to small-scale properties of the galaxy nuclei, rather
than global properties of the entire galaxies.  In particular, for this
reason, source sizes and position angles do not include emission from
the galaxy disks.

\section{Measurements}

In Figure 1 we present the contour plots of the observed galaxies. Of
the 55 galaxies observed, only TOL1238-364 was not detected. The total
flux densities of the sources (Stokes parameter I) were measured
integrating over the images, taking into account the background level,
and are given in column 7 of Table 1. We also used the deconvolved
images to determine the 1$-\sigma$ noise level of the images. These
values vary from 22~$\mu$Jy/beam to 109~$\mu$Jy/beam, but typically are
around 30~$\mu$Jy/beam to 60~$\mu$Jy/beam.  The smaller values
correspond to the galaxies with the higher integration time, while the
higher ones correspond to the galaxies with shorter integrations, or
those observed through a high air mass. These values are given in
column 8 of Table 1.

The radio images were decomposed by fitting gaussians to their
individual components. The positions of these components and their
respective fluxes are given in Table 2. Comparing the total fluxes
obtained in this method with the fluxes measured directly from the
images (Table 1) we find an agreement of 5\% to 10\% for most of the
galaxies, especially the ones with unresolved emission. In the case of
slightly resolved sources and sources with diffuse emission, the
agreement is not always as good as 10\%, since the gaussians tend to
separate the compact emission only.

For those galaxies with extended radio emission, we present in column 9
of Table 1 the measured Position Angle of this emission (PA$_{RAD}$).
These values were obtained from the positions of the individual
gaussians fitted to the data, or, in the case of galaxies with more
extended emission, measured directly on the images. These measurements
have an uncertainty of $3-5^{\circ}$ for linear extended sources and
$5-10^{\circ}$ for slightly resolved ones. The linear sizes of the
radio sources, calculated assuming H$_o=75$km s$^{-1}$ Mpc$^{-1}$, are
given in column 10 of Table 1.  Finally, the structure of the radio
emission, classified according to the Ulvestad \& Wilson (1984b)
definition is given in column 11 of Table 1 (L=linear, D=diffuse,
S=slightly extended, U=unresolved).

\subsection{Flux Density and Position Measurement Errors}

Errors involved in the flux density measurements are due to two
different sources. The first and most important one is the uncertainty
in transferring the amplitude scale of the primary flux calibrator,
which is of the order
of 5\% at 8.46 GHz and dominates the errors. The second source of error
is the noise in the images. These two values have to be added in
quadrature, and the contribution from the noise in the images is
important only for those galaxies where this value is similar to the
one due to the amplitude scale of the calibrator, usually objects with
integrated fluxes smaller than 1.0~mJy.

We estimate that the errors in the positions of the individual
components presented in Table~2 are of the order of
0.01-0.03$^{\prime\prime}$, in both Right Ascention and Declination.
Several sources contribute to this error, and they have to be added in
quadrature. The first one is the uncertainty in the positions of the
phase calibrators. Most of the observations presented here were done
with A-category calibrators, which have position uncertainties
$<0.002^{\prime\prime}$, so this error is negligible in most of the
cases. The second source is the error in the gaussian fitting to the
individual components, which we determine to be of the order of
0.005$^{\prime\prime}$ for the measurements presented here. A third
source of error is due to the strength of the detected emission. For a
detection with a given signal to noise ratio (S/N), the position of the
source cannot be determined with an accuracy better than (beam
size)/(S/N). The typical beamsize of our observations is
0.25$^{\prime\prime}$, so this source of error is dominant for
components where the detection has S/N$\leq$10.

\subsection{Individual Objects}

In this section we discuss the observed radio structures of the galaxies
with extended emission.

\subsubsection{UGC2514}
This galaxy presents a nuclear point source and an extended source to
the SW, along PA=56$^{\circ}$. The total extent of the radio emission
is 70 parsecs.

\subsubsection{IRAS03106-0254}
Norris et al. (1990) detected a compact nuclear source in this galaxy,
with flux of 12 mJy at 2.3 GHz. Our radio image was decomposed into 3
components, extended along the NE-SW direction PA=37$^{\circ}$. The
radio source is extended by 850 parsecs.

\subsubsection{IRAS04502-0317}
Spectropolarimetric observations (Kay 1994) show that this galaxy
presents a very small amount of polarization $\approx$0.5\% around
4400\AA\ and the HST F606W image (Malkan, Gorjian \& Tam 1998) shows a
dust lane crossing the nucleus. This galaxy has a weak double radio
source, extended along  PA=22$^{\circ}$, with a dimension of 110
parsecs.

\subsubsection{IRAS04507+0358}
The stellar population synthesis of the nuclear spectrum of this galaxy
shows that it is dominated by old stars, with a small ($\approx$5\%)
contribution from 100Myr stars (Schmitt, Storchi-Bergmann \& Cid
Fernandes 1999). The [OIII]$\lambda$5007\AA\ image shows halo like
extended emission (Mulchaey, Wilson \& Tsvetanov 1996). The radio
emission of this galaxy is slightly resolved, having an extent of 210
parsecs along PA=153$^{\circ}$.

\subsubsection{MRK6}
The radio image of this galaxy presents an intricate morphology, with a
linear double structure along the N-S direction (PA$=-3^{\circ}$) and a
low surface density structure, similar to a ring, approximately
perpendicular to the jet. This ring structure has been previously
detected by Baum et al.  (1993), Kukula et al. (1996) and Xu et al.
(1997). Baum et al. (1993) also detected radio lobes extending up to
40$^{\prime\prime}$ from the nucleus. The jet is extended by 440
parsecs, while the ring like structure is extended by $\sim$1800
parsecs. Neutral Hydrogen mapping of the nuclear region by Gallimore et
al.  (1999) shows HI absorption against the northern part of the jet,
possible due to a dust lane crossing north of the nucleus. Kukula et
al. (1996) compared MERLIN high resolution radio observations and
[OIII]$\lambda$5007\AA\ images of this galaxy, showing that they are
misaligned. They suggest that this misalignment is due in part to a
projection effect, where the jet is emitted at an angle relative to the
galaxy disk and the gas emission comes from ionized gas in the disk of
the galaxy.

\subsubsection{MRK79}
Ulvestad \& Wilson (1984a) detected extended radio emission at 6~cm
and 20~cm in this galaxy. The 3.6cm radio image presented here confirms
their results, an asymmetric triple radio structure along
PA$=11^{\circ}$. The distance between the nucleus and the northern
source is 800 parsecs, while that to the southern source is 460
parsecs. Nazarova, O'Brien \& Ward (1996) detected extended line
emission along PA$=12^{\circ}$, based on spectroscopic observations.

\subsubsection{MCG-01-24-012}
The radio emission of this galaxy is extended to the W (PA$=89^{\circ}$)
by 130 parsecs. 

\subsubsection{NGC3393}
The X-ray spectrum of this galaxy (Maiolino et al. 1998) has a hard
flux F$_{2-10Kev}=3.9\times10^{-13}$ erg cm$^{-2}$ s$^{-1}$, and can be
fitted with a power law Compton thick absorbed spectrum (N$_{H}>10^{25}$
cm$^{-2}$) reflected by cold gas in our direction. A detailed analysis
of the ultraviolet spectrum of this galaxy can be found in D\'{\i}az,
Prieto \& Wamsteker (1988).  The radio emission can be decomposed into
4 different components along the NE-SW direction, with total extent of
680 parsecs. The nucleus and the stronger components are along
PA$=56^{\circ}$. The SW component shows some diffuse emission to the S,
and the NE component shows a second component to the N. A comparison
between the radio image and the HST [OIII]$\lambda$5007\AA\ emission
(Schmitt \& Kinney 1996), shows that the radio emission is encircled by
the [OIII] emission, suggesting a shock front that ionizes the gas.

\subsubsection{IRAS11215-2806}
The nuclear stellar population of this galaxy (Schmitt et al. 1999) is
dominated by old stars, with a small contribution ($\approx$5\%) from
100Myr old stars. This galaxy shows several radio blobs. The strongest
ones are extended by $\sim$200 parsecs along PA$=75^{\circ}$, with a
weaker source 400 parsecs to the SW of the nucleus (PA$=226^{\circ}$).

\subsubsection{MCG-05-27-013}
The nuclear stellar population of this galaxy (Schmitt et al. 1999)
has contributions only from stars older than 1~Gyr. The HST F606W image
(Malkan et al. 1998) shows a strong dust lane to the N of the nucleus.
The overall radio emission of this galaxy is extended by 1530 parsecs
along PA$=2^{\circ}$. It shows the nucleus, a detached blob to the N
and a diffuse source to the S.

\subsubsection{MRK176}
This is an interacting Seyfert 2 galaxy. According
to Veilleux, Goodrich \& Hill (1997), it may have a broad
Pa$\beta$ emission line.  The radio emission is slightly
extended by 140 parsecs along PA$=92^{\circ}$.

\subsubsection{MCG-04-31-030}
According to Pogge (1989) the [OIII] and H$\alpha$ images of this
galaxy do not show extended emission. The hard X-ray spectrum can be
fitted with a double power-law and has a flux
F$_{2-10~Kev}$=3.56$\times10^{-12}$erg cm$^{-2}$ s$^{-1}$ (Turner et
al. 1997). The HST F606W image (Malkan et al. 1998) shows a bright
nuclear source and a dust lane to the N of the nucleus. This galaxy
shows a low flux density linear structure extended by 480 parsecs along
PA$=85^{\circ}$.

\subsubsection{MCG-03-34-064}
According to Roy et al. (1998), this galaxy has a compact nuclear radio
source with a flux of 26~mJy at 2.3~GHz. The optical and ultraviolet
spectra show narrow emission lines with blueward assymmetries
and possibly a broad CIV$\lambda$1550\AA\ line (De Robertis, Hutchings
\& Pitts 1988). The radio emission is extended by 280 parsecs. It
starts as a linear structure to the SW (PA$=219^{\circ}$) and after
$\sim$100 parsecs it bends to the S, along PA$\approx180^{\circ}$. Kay
(1994) spectropolarimetric data showed that the nuclear spectrum of
this galaxy is polarized by 7.4\% at 4400\AA, along PA$=109^{\circ}$,
approximately perpendicular to the extended radio emission.
We adopt the innermost PA as the direction of the jet.

\subsubsection{ESO383-G18}
This galaxy shows a slightly extended radio structure to the N of the
nucleus (PA$=-2^{\circ}$) extended by 110 parsecs.

\subsubsection{IRAS14082+1347}
The Malkan et al. (1998) HST F606W image of this galaxy shows a dust lane
crossing to the north of the nucleus. The radio emission is slightly
extended to the W, along PA$=279^{\circ}$, with a total extent of 55
parsecs.

\subsubsection{NGC5506}
According to Braatz, Wilson \& Henkel (1996) this galaxy has an H$_2$O
maser. The radio emission shows a linear structure along the E-W
direction, surrounded by diffuse emission, with a total extension of
300 parsecs. The VLBA image from Roy et al. (2000) shows a double radio
source along PA$=70^{\circ}$, which will be used by Kinney et al.
(2000). This galaxy is almost edge-on and the HI image (Gallimore et al.
1999) shows absorption against the nuclear source.  The hard X-ray
spectrum has a flux F$_{2-10Kev}=8.38\times10^{-11}$ erg cm$^{-2}$
s$^{-1}$ and is absorbed by a column density N$_{H}=3.4\times10^{22}$
cm$^{-2}$ (Bassani et al.  1999). Maiolino et al. (1994) detected
double peaked line profiles, which they interpreted as outflow
along PA$=-16^{\circ}$.  Colbert et al. (1998) detected extended soft
X-ray emission along the same direction, PA$=-20^{\circ}$, which is
also coincident with the extended radio emission.

\subsubsection{IRAS14317-3237}
This galaxy shows a radio structure extended by 290 parsec along
PA$=169^{\circ}$.

\subsubsection{UGC9944}
The radio emission of this galaxy is a symmetric triple source with
total extent of 3430 parsecs along PA$=67^{\circ}$. The SW component
can be decomposed into 2 sources, separated by $\approx$140 parsecs.

\subsubsection{AKN564}
According to Boller, Brandt \& Fink (1996) this is an X-ray variable
Narrow Line Seyfert 1 galaxy with a steep spectrum, photon index
$\Gamma\approx3$ in the range 0.1-2.4Kev. It has a triple radio source
along the N-S direction (PA$=6^{\circ}$), extended by 320 parsecs.
This is consistent with the result shown by Moran (2000).

\subsubsection{MCG+8-11-11}
The first radio observations of this galaxy were done by Wilson \&
Willis (1981), using the VLA at 6cm, which showed a strong nuclear
source and some extended emission in the N-S direction. Ulvestad \&
Wilson (1986) presented a VLA image at 2cm, showing a
triple radio source, extended by
$\approx1^{\prime\prime}$ along PA$=127^{\circ}$. Our radio image at
3.6cm shows a central component extended along PA$=128^{\circ}$,
surrounded by diffuse emission along the N-S direction, with a total
extent of 1230 parsecs. This structure is similar to the one observed
by Ulvestad \& Wilson (1986) at 6cm.  The HST [OIII] and H$\alpha$
images of this galaxy show no evidence of extended emission (Bower et
al. 1994; Schmitt \& Kinney 1996).

\subsubsection{NGC4941}
Maiolino et al. (1999) presented the X-ray spectrum of this galaxy,
which has a soft band flux F$_{0.5-2Kev}=2.08\times10^{-13}$ erg
cm$^{-2}$ s$^{-1}$, a hard band flux F$_{2-10Kev}=6.63\times10^{-13}$
erg cm$^{-2}$ s$^{-1}$, and can be fitted with a power law absorbed by
a column density N$_{H}=4.5\times10^{23}$ cm$^{-2}$. The [OIII]
emission (Pogge 1989) has a halo like morphology and the H$\alpha$
image shows HII regions along the spiral arms. This is the smallest
extended source in our sample, presenting a double radio source
extended by 15 parsecs along PA$=-25^{\circ}$. This source was only
slightly resolved in previous radio observations (Ulvestad \& Wilson
1989).

\section{Summary}

We presented VLA A-configuration observations of 55 Seyfert galaxies at
8.46GHz. These galaxies are part of a sample of 88 Seyfert galaxies
selected from a mostly isotropic property, the flux at 60$\mu$m, and
warm infrared colors. Only one of the 55 galaxies, TOL1238-364 was not
detected, and 21 galaxies show extended emission. We measured the radio
flux densities of these galaxies, the sizes of their emitting regions,
the Position Angle of the extended radio emission (PA$_{RAD}$) and their
morphologies. These measurements have been combined with information from
broad-band B and I observations to study the orientation of radio jets
relative to the host galaxy disk (Kinney et al. 2000).

\acknowledgements 

We would like to thank Jim Pringle for useful discussions.
HRS would like to thank the hospitality of the NRAO staff in Socorro,
where the data was reduced and analyzed.
This work was supported by NASA grants NAGW-3757, AR-5810.01-94A,
AR-6389.01-94A and the HST Director Discretionary fund D0001.82223.
This research made use of the NASA/IPAC Extragalactic Database (NED),
which is operated by the Jep Propulsion Laboratory, Caltech, under
contract with NASA. The National Radio Astronomy Observatory is a
facility of the National Science Foundation operated under cooperative
agreement by Associsted Universities, Inc.

\begin{figure}
\caption{Radio continuum 8.46 GHz (3.6cm) contour plots presented
following the order given in Table 1. TOL1238-364 is not presented here
because it was not detected. The contours start at 3$\sigma$ above the
background level and increase in powers of $\sqrt2$
(3$\sigma\times2^{(n/2)}$, n=0,1,2,...). We also plot, as a dashed
line, the contour plots corresponding to the $3\sigma$ level.
The lower left corner of each panel shows the shape and size
of the restoring beam.}
\end{figure}

\end{document}


\begin{small}
\begin{deluxetable}{rlcrclrrrrll}
\tablewidth{0pc}
\tablecaption{Observations and Measurements}
\tablehead{\colhead{Number}&\colhead{Name}&\colhead{Type}&\colhead{Date}
&\colhead{VLA proposal}&\colhead{Observing}&\colhead{F$_{3.6cm}$}&\colhead{$\sigma$}
&\colhead{PA$_{RAD}$}&\colhead{Size}&\colhead{Morph.}&\colhead{Ref.}\cr
&&&&&Method&(mJy)&($\mu$Jy/B)&(degrees)&(pc)&&}
\startdata
26 &MRK\,1              &2&4-12-98 &AA226 &CSC...&15.4  &30  &--  &$<$30 &U    &\cr
37 &IRAS\,01475$-$0740  &2&7-15-95 &AK394 &CS    &135.4 &62  &--  &$<$35 &U    &1\cr
47 &MRK\,1040           &1&7-15-95 &AK394 &CS    &1.2   &89  &--  &$<$30 &U    &1\cr
53 &UGC\,2024           &2&4-12-98 &AA226 &CSC...&1.2   &31  &--  &$<$45 &U    &\cr
67 &MCG$-$02$-$08$-$039 &2&7-15-95 &AK394 &CS    &2.0   &62  &--  &$<$55 &U    &1\cr
68 &UGC\,2514           &1&4-12-98 &AA226 &CSC...&0.6   &31  &56  &70    &L    &\cr
75 &IRAS\,03106$-$0254  &2&5-3-98  &AA226 &CSC...&11.3  &29  &37  &850   &L    &\cr
78 &IRAS\,03125$+$0119  &2&5-3-98  &AA226 &CSC...&9.1   &31  &--  &$<$45 &U    &\cr
141&IRAS\, 04385$-$0828	&2&7-16-95 &AK394 &CS    &7.6   &53  &--  &$<$30 &U    &1\cr
154&IRAS\, 04502$-$0317	&2&5-3-98  &AA226 &CSC...&0.5   &32  &22  &110   &L    &\cr
156&IRAS\, 04507$+$0358	&2&5-13-98 &AA226 &CSC...&0.9   &33  &153 &210   &S    &\cr
174&MCG$-$05$-$13$-$017 &1&7-15-95 &AK394 &CS    &2.5   &46  &--  &$<$25 &U    &1\cr
203&UGC\,3478           &1&4-13-98 &AA226 &CSC...&1.6   &32  &--  &$<$25 &U    &\cr
209&MRK\,6              &1&7-20-95 &AK407 &CSC...&39.5  &24  &-3  &440   &L    &1,2\cr
225&MRK\,79             &1&7-15-95 &AK394 &CS    &2.9   &40  &11  &1260  &L    &1\cr
227&MRK\,10             &1&4-13-98 &AA226 &CSC...&0.3   &31  &--  &$<$55 &U    &\cr
233&UGC\,4155           &1&4-13-98 &AA226 &CSC...&2.4   &31  &--  &$<$50 &U    &\cr
253&MCG$-$01$-$24$-$012	&2&4-13-98 &AA226 &CSC...&9.1   &32  &89  &140   &L    &\cr
272&NGC\,3393           &2&11-29-92&AB618 &CSC...&13.0  &24  &56  &680   &L    &\cr
281&IRAS\,11215$-$2806  &2&4-13-98 &AA226 &CSC...&12.0  &35  &75  &400   &L    &\cr
282&MCG$-$05$-$27$-$013	&2&4-13-98 &AA226 &CSC...&8.4   &31  &2   &1530  &L    &\cr
283&MRK\,176            &2&4-13-98 &AA226 &CSC...&6.3   &29  &92  &140   &S    &\cr
286&NGC\,3783           &1&9-1-91  &AW278 &CSC...&8.0   &49  &--  &$<$15 &U    &\cr
301&NGC\,4593           &1&4-13-98 &AA226 &CSC...&2.1   &31  &--  &$<$15 &U    &\cr
302&TOL\,1238-364       &2&7-16-95 &AK394 &CS    &--    &46  &--  &--    &--   &1\cr
306&NGC\,4704           &2&4-13-98 &AA226 &CSC...&0.9   &31  &--  &$<$55 &U    &\cr
309&MCG$-$02$-$33$-$034	&1&7-16-95 &AK394 &CS    &1.3   &49  &--  &$<$30 &U    &1\cr
313&MCG$-$04$-$31$-$030	&2&7-16-95 &AK394 &CS    &9.4   &49  &85  &480   &L    &1\cr
314&IRAS\,13059$-$2407 	&2&3-8-90  &AD244 &CS    &126.0 &60  &--  &$<$30 &U    &\cr
317&MCG$-$03$-$34$-$064	&2&7-16-95 &AK394 &CS    &49.0  &67  &39  &280   &L    &1\cr
322&ESO\,383$-$G18      &2&4-13-98 &AA226 &CSC...&2.1   &33  &178 &110   &S    &\cr
329&NGC\,5347           &2&4-13-98 &AA226 &CSC...&1.6   &30  &--  &$<$15 &U    &\cr
340&IRAS\,14082$+$1347  &2&4-13-98 &AA226 &CSC...&3.5   &30  &99  &55    &S    &\cr
341&NGC\,5506           &2&9-1-91  &AW278 &CSC...&104.0  &43  &70$^a$&300 &L$+$D&\cr
349&IRAS\,14317$-$3237  &2&4-13-98 &AA226 &CSC...&2.2   &33  &169 &290   &L    &\cr
354&IRAS\,14434$+$2714  &2&4-13-98 &AA226 &CSC...&0.4   &31  &--  &$<$120&U    &\cr
369&UGC\,9826           &1&4-13-98 &AA226 &CSC...&0.2   &31  &--  &$<$55 &U    &\cr
377&UGC\,9944           &2&4-13-98 &AA226 &CSC...&6.3   &32  &67  &3430  &L    &\cr
383&IRAS\,15480$-$0344  &2&7-16-95 &AK394 &CS    &11.2  &40  &--  &$<$60 &U    &1\cr
409&IRAS\,16288$+$3929  &2&4-13-98 &AA226 &CSC...&5.1   &30  &--  &$<$60 &U    &\cr
418&IRAS\,16382$-$0613  &2&4-13-98 &AA226 &CSC...&2.7   &32  &--  &$<$55 &U    &\cr
445&UGC\,10889          &2&4-13-98 &AA226 &CSC...&1.1   &32  &--  &$<$55 &U    &\cr
447&MCG$+$03$-$45$-$003	&2&4-13-98 &AA226 &CSC...&0.3   &32  &--  &$<$50 &U    &\cr
510&UGC\,11630          &2&4-13-98 &AA226 &CSC...&0.5   &31  &--  &$<$25 &U    &\cr
530&NGC\,7213           &1&7-16-95 &AK394 &CS    &188.0 &109 &--  &$<$15 &U    &1\cr
537&MRK\,915            &1&4-13-98 &AA226 &CSC...&15.8  &32  &--  &$<$45 &U    &\cr
540&AKN\,564            &1&4-13-98 &AA226 &CSC...&7.0   &27  &6   &320   &L    &\cr
549&UGC\,12348          &2&4-13-98 &AA226 &CSC...&1.6   &31  &--  &$<$120&U    &\cr
590&MRK\,590            &1&4-12-98 &AA226 &CSC...&3.4   &30  &--  &$<$50 &U    &\cr
594&MRK\,1058           &2&5-13-98 &AA226 &CSC...&0.2   &35  &--  &$<$35 &U    &\cr
615&MCG\,+8-11-11       &1&5-13-98 &AA226 &CSC...&38.4  &30  &128 &1230  &L$+$D&\cr
627&MRK\,705            &1&3-1-94  &AK350 &CSC...&2.0   &50  &--  &$<$55 &U    &\cr
634&NGC\,3281           &2&3-24-94 &AP276 &CSC...&17.0  &22  &--  &$<$60 &U    &\cr
665&NGC\,4941           &2&7-16-95 &AK394 &CS    &4.1   &45  &-25 &15    &L    &1\cr
703&UGC\,10683B         &1&4-13-98 &AA226 &CSC...&0.6   &32  &--  &$<$60 &U    &\cr
\tablenotetext{}{Col. 1 -- Numbers of the galaxies in the de Grijp et al. (1987) catalog.}
\tablenotetext{}{Col. 2 -- Names of the galaxies.}
\tablenotetext{}{Col. 3 -- Activity types.}
\tablenotetext{}{Col. 4 -- Dates when the observation were done.}
\tablenotetext{}{Col. 5 -- VLA proposals from which the data were obtained.}
\tablenotetext{}{Col. 6 -- Method used to observe the source (S) and the phase calibrator (C).}
\tablenotetext{}{Col. 7 -- Integrated 3.6cm flux in mJy.}
\tablenotetext{}{Col. 8 -- 1-$\sigma$ noise level in $\mu$Jy/Beam.}
\tablenotetext{}{Col. 9 -- Measured position angle of the extended radio emission.}
\tablenotetext{}{Col. 10 -- Size of the radio emitting region (calculated
assuming H$_0=75$ km s$^{-1}$ Mpc$^{-1}$).}
\tablenotetext{}{Col. 11 -- Morphology of the radio emission, U=unresolved,
S=slightly resolved, L=Linear, D=diffuse.}
\tablenotetext{}{Col. 12 -- For data obtained from the archive, the reference where it
has been published: 1) Thean et al. (2000), 2) Xu et al. (1997).}
\tablenotetext{a}{PA$_{RAD}$ obtained from VLBA measurements (Roy et al. 2000).}
\enddata
\end{deluxetable}
\end{small}

\clearpage
\newpage

\begin{small}
\begin{deluxetable}{lrrr}
\tablewidth{0pc}
\tablecaption{Positions and Fluxes of Individual Components}
\tablehead{\colhead{Name}&\colhead{RA(J2000)}&\colhead{Decl.(J2000)}&
\colhead{F$_{3.6cm}$}\cr
&&&(mJy)}
\startdata
MRK\,1              &01 16 07.215&  +33 05 21.60&  14.8  \cr
IRAS\,01475$-$0740  &01 50 02.696&  -07 25 48.53& 135.4  \cr
MRK\,1040           &02 28 14.476&  +31 18 41.49&   1.1  \cr
UGC\,2024           &02 33 01.240&  +00 25 14.89&   0.7  \cr
MCG$-$02$-$08$-$039 &03 00 30.621&  -11 24 57.10&   2.0  \cr
UGC\,2514           &03 03 49.091&  -01 06 13.64&   0.2  \cr
                    &03 03 49.109&  -01 06 13.47&   0.4  \cr
IRAS\,03106$-$0254  &03 13 08.753&  -02 43 20.02&   1.0  \cr
                    &03 13 08.780&  -02 43 19.27&   5.8  \cr
                    &03 13 08.786&  -02 43 19.14&   4.0  \cr
IRAS\,03125$+$0119  &03 15 05.341&  +01 30 30.16&   8.8  \cr
IRAS\, 04385$-$0828 &04 40 54.964&  -08 22 22.09&   7.5  \cr
IRAS\, 04502$-$0317 &04 52 44.532&  -03 12 57.46&   0.2  \cr
                    &04 52 44.541&  -03 12 57.14&   0.2  \cr
IRAS\, 04507$+$0358 &04 53 25.742&  +04 03 41.83&   0.6  \cr
                    &04 53 25.753&  +04 03 41.51&   0.2  \cr
MCG$-$05$-$13$-$017 &05 19 35.815&  -32 39 27.81&   2.5  \cr
UGC\,3478           &06 32 47.172&  +63 40 25.32&   1.4  \cr
MRK\,6              &06 52 12.400&  +74 25 38.04&  10.4  \cr
                    &06 52 12.410&  +74 25 37.03&  17.0  \cr
                    &06 52 12.400&  +74 25 36.81&   9.9  \cr
MRK\,79             &07 42 32.785&  +49 48 33.85&   0.7  \cr
                    &07 42 32.810&  +49 48 34.90&   1.3  \cr
                    &07 42 32.848&  +49 48 36.70&   0.9  \cr
MRK\,10             &07 47 29.084&  +60 56 00.62&   0.3  \cr
UGC\,4155           &08 00 20.980&  +26 36 48.62&   2.4  \cr
MCG$-$01$-$24$-$012 &09 20 46.236&  -08 03 21.96&   0.9  \cr
                    &09 20 46.259&  -08 03 21.95&   8.0  \cr
NGC\,3393           &10 48 23.380&  -25 09 44.07&   8.4  \cr
                    &10 48 23.440&  -25 09 43.51&   0.7  \cr
                    &10 48 23.500&  -25 09 42.98&   0.8  \cr
                    &10 48 23.520&  -25 09 42.37&   1.4  \cr
IRAS\,11215$-$2806  &11 24 02.641&  -28 23 16.45&   0.7  \cr
                    &11 24 02.696&  -28 23 15.54&   1.7  \cr
                    &11 24 02.714&  -28 23 15.48&   4.0  \cr
                    &11 24 02.729&  -28 23 15.42&   4.4  \cr
MCG$-$05$-$27$-$013 &11 27 23.437&  -29 15 27.78&   3.7  \cr
                    &11 27 23.441&  -29 15 29.21&   0.5  \cr
                    &11 27 23.448&  -29 15 25.90&   0.5  \cr
MRK\,176            &11 32 40.238&  +52 57 01.31&   6.3  \cr
NGC\,3783           &11 39 01.670&  -37 44 18.85&   7.7  \cr
NGC\,4593           &12 39 39.444&  -05 20 39.04&   2.1  \cr
NGC\,4704           &12 48 46.427&  +41 55 16.53&   0.8  \cr
MCG$-$02$-$33$-$034 &12 52 12.475&  -13 24 53.38&   1.2  \cr
MCG$-$04$-$31$-$030 &13 07 05.929&  -23 40 36.46&   5.7  \cr
                    &13 07 05.906&  -23 40 36.49&   3.5  \cr
IRAS\,13059$-$2407  &13 08 41.970&  -24 22 57.99& 126.0  \cr
MCG$-$03$-$34$-$064 &13 22 24.471&  -16 43 42.35&  31.6  \cr
                    &13 22 24.456&  -16 43 42.61&  15.5  \cr
ESO\,383$-$G18      &13 33 26.105&  -34 00 53.25&   1.7  \cr
NGC\,5347           &13 53 17.796&  +33 29 26.94&   1.6  \cr
IRAS\,14082$+$1347  &14 10 41.350&  +13 33 28.87&   3.2  \cr
NGC\,5506           &14 13 14.830&  -03 12 27.76&  73.6  \cr
IRAS\,14317$-$3237  &14 34 45.474&  -32 50 32.29&   1.1  \cr
                    &14 34 45.482&  -32 50 32.86&   0.8  \cr
IRAS\,14434$+$2714  &14 45 36.846&  +27 02 05.58&   0.4$^a$\cr
UGC\,9826           &15 21 32.983&  +39 12 02.72&   0.2  \cr
UGC\,9944           &15 35 46.919&  +73 27 01.07&   1.9  \cr
                    &15 35 46.957&  +73 27 01.31&   1.7  \cr
                    &15 35 47.680&  +73 27 02.50&   0.6  \cr
                    &15 35 48.443&  +73 27 03.82&   1.7  \cr
IRAS\,15480$-$0344  &15 50 41.498&  -03 53 18.05&  11.2  \cr
IRAS\,16288$+$3929  &16 30 32.656&  +39 23 03.14&   4.8  \cr
IRAS\,16382$-$0613  &16 40 52.248&  -06 18 52.40&   2.1  \cr
UGC\,10889          &17 30 21.410&  +59 38 23.66&   1.1  \cr
MCG$+$03$-$45$-$003 &17 35 33.761&  +20 47 47.04&   0.3  \cr
UGC\,11630          &20 47 34.096&  +00 24 41.76&   0.4  \cr
NGC\,7213           &22 09 16.200&  -47 09 58.85& 187.9  \cr
MRK\,915            &22 36 46.503&  -12 32 42.80&  14.5  \cr
                    &22 36 46.467&  -12 32 43.36&   0.2  \tablebreak
AKN\,564            &22 42 39.334&  +29 43 31.59&   0.9  \cr
                    &22 42 39.337&  +29 43 31.24&   3.0  \cr
                    &22 42 39.340&  +29 43 30.91&   3.1  \cr
UGC\,12348          &23 05 18.836&  +00 11 22.16&   1.1  \cr
MRK\,590            &02 14 33.564&  -00 46 00.37&   3.2  \cr
MRK\,1058           &02 49 51.842&  +34 59 16.90&   0.2  \cr
MCG\,+8-11-11       &05 54 53.613&  +46 26 21.57&  24.4$^b$\cr
MRK\,705            &09 26 03.260&  +12 44 03.46&   2.0  \cr
NGC\,3281           &10 31 52.050&  -34 51 13.55&  17.0  \cr
NGC\,4941           &13 04 13.110&  -05 33 05.96&   1.2  \cr
                    &13 04 13.104&  -05 33 05.77&   2.8  \cr
UGC\,10683B         &17 05 00.393&  -01 32 28.53&   0.6  \cr
\tablenotetext{a}{There is a second radio source 21$^{\prime\prime}$~N
of IRAS14434+2714, probably related to a fainter background galaxy, which
can be seen on the HST archival image. This source can be decomposed
in two components at RA(J2000)=14:45:36.909,  DEC(J2000)=+27:02:27.11
and RA(J2000)=14:45:36.915,  DEC(J2000)=+27:02:27.30 with 3.6cm fluxes
0.9mJy and 7.5mJy, respectively.}
\tablenotetext{b}{Core component only.}
\enddata
\end{deluxetable}
\end{small}